\documentclass[twocolumn]{article}

\usepackage[applemac]{inputenc}
\usepackage[T1]{fontenc} 
\usepackage[english]{babel}
\usepackage{amsmath,amssymb,amsthm}
\usepackage{natbib}
\usepackage{setspace}
\usepackage{graphicx,tikz}
\usepackage{array,multirow}
\usepackage{float}

\title{Inter-Rater Reliability is Individual Fairness}

\author{Tim Räz\footnote{University of Bern, Institute of Philosophy, L\"anggassstrasse 49a,  3012 Bern, Switzerland. e-mail: tim.raez@posteo.de}}

\theoremstyle{definition}

\newtheorem{proposition}{Proposition}

\begin{document}

\maketitle

\begin{abstract}
In this note, a connection between inter-rater reliability and individual fairness is established. It is shown that inter-rater reliability is a special case of individual fairness, a notion of fairness requiring that similar people are treated similarly. 
\end{abstract}

\section{Introduction}

In this note, a connection between inter-rater reliability (IRR) and individual fairness is established. IRR is the degree to which individuals subjected to a predictive model receive consistent predictions by different raters. IRR is an important property of predictive models, because it a necessary condition of predictive validity, the degree to which predictions are correct. Here it is shown that IRR is equivalent to a special case of individual fairness (IF), a fairness notion that requires similar people to be treated similarly. 

Informally, the idea is to let the similarity requirement of IF be the identity relation for individuals, such that each individual is always supposed to get the same prediction. The extent to which this condition is violated is exactly what IRR measures. This equivalence implies that whenever we examine inter-rater reliability, we also examine the degree to which individual fairness is satisfied. It also means that differences in reliability between socially salient groups can be interpreted as differences in individual fairness, aggregated for these groups.

\section{Background}

\subsection{Individual Fairness (IF)}

Different notions of fairness are discussed in the debate on fair-ML \citep{baroc2019}. Notions of group fairness are concerned with the distribution of predictions and ground truth for different, socially salient groups. The present paper does not deal with traditional notions of group fairness, but it provides a novel interpretation of reliability as individual fairness on the aggregate level.

Informally speaking, a predictor satisfies individual fairness (IF) if individuals with similar features get similar predictions. This idea has been formalized by \citet{dwork2012}: A (possibly randomized) predictor $f: X \rightarrow \Delta(Y)$ should satisfy a so-called Lipschitz condition. Note that here, only non-randomized predictors $f:X \rightarrow Y$ will be considered. Formally speaking, given a suitable metric\footnote{Strictly speaking, a pseudo metric, i.e., a function $d: X \times X \rightarrow \mathbb{R}$ with $d(x,y) \geq 0$, $d(x,y)=d(y,x)$ and $d(x,x) = 0$, cf. \citet{dwork2012}.} $d$ between individuals $x, x' \in X$ and a suitable metric $D$ between predictions $f(x)$ and $f(x')$ for these individuals, the predictor $f$ satisfies $(d, D)$-individual fairness if $D\big(f(x),f(x')\big) \leq d(x,x')$. This means that the distance $d$ between individuals should limit the distance $D$ of predictions for these individuals. In its intuitive formulation, individual fairness has a long history, which has been traced back to Aristotle's requirement of consistency for judgements, \citep{binns2020}.

\subsection{Inter-rater Reliability (IRR)}

Inter-rater reliability is an important property of predictive models. It is measured because the inputs of a predictive model may contain errors, which in turn may affect predictions. For example, in recidivism risk assessment \citep{desma2018}, a risk assessment instrument (RAI) is a function $f:X \rightarrow \hat{Y}$ that takes features of individuals $x \in X$ as inputs and assigns them predictions $f(x) = \hat{y} \in \hat{Y}$ in the form of scores (e.g. probabilities of reoffending), or risk categories (``will reoffend'', ``will not reoffend''). Before predictions can be made, the features of individuals $i \in I$ have to be determined through a rating process. Ratings are determined by human raters who fill out a questionnaire on the basis of interviews with the individual, case files, and so on. The rating can be represented by a function  $x: I \rightarrow X$, $i \mapsto x(i)$. The instrument $f$ then takes ratings of individuals $x(i) \in X$ as inputs. 

The rating $x(i)$ of individual $i$ should represent $i$ accurately. In particular, $x(i)$ should not vary from rater to rater. Inter-rater reliability (IRR) measures the extent to which this is satisfied. To measure IRR, a set of individuals $I$ has to be rated by at least two raters, $r$ and $s$. Here the focus is on the reliability of predictions, that is, the extent to which ratings $x_r(i)$ and $x_s(i)$ of individual $i$ by raters $r$ and $s$ lead to the same or different predictions $f(x_r(i))$ and $f(x_s(i))$. To calculate IRR of predictions, one needs data of the following form:

\begin{table}[h]
\begin{center}
\begin{tabular}{|c|c|c|}
\hline
individual & rater r & rater s \\
\hline
1 & $f(x_r(1))$ & $f(x_s(1))$ \\
2 & $f(x_r(2))$ & $f(x_s(2))$ \\
... & ... & ... \\
n & $f(x_r(n))$ & $f(x_s(n))$ \\
\hline
\end{tabular}
\end{center}
\caption{Data for inter-rater reliability of predictions.}
\label{irr_predictions}
\end{table}

There are different IRR statistics depending on the kind of prediction. If $f$ makes binary predictions, i.e., $f(x_r(i)), f(x_s(i)) \in \{0, 1\}$, one can form a confusion matrix on the basis of table \ref{irr_predictions}, and calculate, say, Cohen's Kappa \citep{cohen1960}. If $f$ makes continuous predictions, such as a score, one can calculate an appropriate intraclass correlation coefficient (ICC) \citep{lilje2019}. These statistics provide methods to evaluate IRR, aggregated from table \ref{irr_predictions}. The usual motivation for measuring (inter-rater) reliability is its relation to validity. Reliability is a necessary condition of validity \citep{desma2018}: a prediction $f(x(i))$ on the basis of a rating $x(i)$ with respect to individual $i$ is only accurate to the extent to which the rating $x(i)$ represents $i$ accurately.

\section{Related Work}

A search in the proceedings of the FAccT conference (up to and including 2022) was conducted to determine whether (inter-rater) reliability has been identified as an issue of fairness. \citet{geige2020a} investigated whether empirical papers using human annotators followed best practices, including the reporting of annotator reliability with respect to class labeling. Note that this is the reliability of labels (ground truth), not the reliability with respect to ratings or predictions. The issue of noise in the labels as a fairness issue is investigated empirically by \citet{wang2021}, who consider the question of how label noise that is correlated with both label class and protected attribute (group membership) can affect both accuracy and (group) fairness. The issue of noise in group membership labels, or inference of group membership, is investigated by \citet{ghazi2022, rieke2022}. The idea that reliability is a fairness issue is stressed in \citet{jacob2021}, where the concept of ``construct reliability'' is proposed: similar inputs to a measurement model (operationalization of a theoretical construct) should lead to similar outputs, or results of measurement. This is very similar to what is proposed here. However, Jacobs et al. explicitly exclude inter-rate reliability from their considerations, because they consider it to be an issue for measurement with ``qualitative methods'' (Ibid., fn. 4).

\section{Reliability is Individual Fairness}

To show that inter-rater reliability is individual fairness, two choices have to be made. First, individual fairness is measured with respect to individuals, not with respect to features of individuals. The metric $d(x,x')$ in the definition of individual fairness is defined on $X \times X$, i.e., between features (ratings) of individuals. Here the metric $d$ on $I \times I$ is considered. This opens up the possibility that the very same individual $i$ is assigned different features if two raters $r$ and $s$ provide different ratings $x_r(i) \neq x_s(i)$. Second, the discrete metric on $I \times I$ is chosen, i.e., $d(i, i') = 0$ if $i = i'$, $d(i, i') = 1$ if $i \neq i'$. This means that we choose the similarity relation to be identity: one and the same individual should get the same prediction. It is now straightforward to prove:

\begin{proposition}
Let $I$ be the set of individuals, $X$ the associated feature space, $\hat{Y}$ the prediction space, $x_r, x_s:I \rightarrow X$ the rating functions of raters $r, s$, and $f:X \rightarrow \hat{Y}$ the predictor function. If the metric $d:I \times I \rightarrow \mathbb{R}$ is discrete, and the metric $D$ is normalized, i.e., $D:\hat{Y} \times \hat{Y} \rightarrow [0, 1]$, then the $(d, D)$-individual fairness of $(f \circ x): I \rightarrow \hat{Y}$ is violated with respect to individual $i$ iff. the predictions for $i$ with respect to the ratings $x_r(i)$ and $x_s(i)$ are different.
\end{proposition}

\begin{proof}
Assume that IF is violated, i.e., $D\big(f(x_r(i)), f(x_s(i'))\big) > d(i,i')$. This is the case iff. $d(i, i') = 0$ and $D\big(f(x_r(i)), f(x_s(i'))\big) > 0$, because $d$ is discrete and $D \leq 1$. $d(i, i') = 0$ holds iff. $i = i'$, because $d$ is discrete. Thus $D\big(f(x_r(i)), f(x_s(i'))\big) > 0$ holds iff. $f(x_r(i)) \neq f(x_s(i))$, i.e., the predictions with respect to the two ratings $r$ and $s$ of one individual are different.
\end{proof}

Note that the predictions are different only if the two ratings are different as well. It is possible that the predictions for different ratings are identical. The assumption that $D$ is normalized excludes violations of IF for different individuals and yields the equivalence of IF and IRR. If IF is supposed to require fairness between different individuals as well, then this assumption can be dropped. Thus, IF encompasses IRR, and is identical to it in the special case of discrete $d$ and normalized $D$.\footnote{Note that IF with respect to individuals as proposed here is similar to, yet distinct from, absolute individual fairness \citep{raez2022}. For absolute individual fairness, the metric $d$ is defined on ground truth $Y$ for individuals.}

\section{Discussion}

The above proposition is formulated for individuals, while IRR is usually calculated on the sample level. Thus, IRR quantifies how much IF is violated in aggregate, and we can make use of the methodology of IRR to measure the degree to which a predictor satisfies IF. In particular, if a predictor is binary, we can make use of IRR statistics for binary predictions, such as Cohen's Kappa, while statistics such as ICCs are appropriate for continuous predictions.

Inter-rater reliability is an important but limited consistency requirement for representing individuals. It is important because it is necessary for adequacy of representation: An individual $i$ that is assigned different features by different raters, $x_r(i) \neq x_s(i)$, is not adequately represented by at least one of them. It is limited because consistency of representation is not sufficient for adequacy of representation: An individual $i$ may be represented inadequately even if two raters agree on an (inadequate) representation. Also, an inadequate representation $x(i)$ does not necessarily lead to an incorrect prediction $f(x(i))$, if this prediction agrees with the prediction $f(x_t(i))$ of the ``true'' representation $x_t(i)$. A rating mistake may have no ramifications and not make a difference for predictions. Measuring violations of reliability of predictions means measuring inadequate representations that make a difference.

Usually, IRR is measured with respect to all individuals in a sample. The connection of IRR with fairness suggests that we could also consider IRR with respect to particular, socially salient groups. Is there a difference in reliability (and thus individual fairness) with respect to, say, race or gender? This important question, which has not received a lot of attention so far, is investigated in a companion paper \citep{raez2023b}.

\end{document}